\documentclass[lettersize,journal]{IEEEtran}
\usepackage{amsmath,amsfonts}
\usepackage{algorithm}
\usepackage{algorithmic}
\usepackage{array}
\usepackage[caption=false,font=normalsize,labelfont=sf,textfont=sf]{subfig}
\usepackage{textcomp}
\usepackage{stfloats}
\usepackage{url}
\usepackage{verbatim}
\usepackage{graphicx}
\usepackage{cite}
\usepackage{diagbox}
\usepackage{amsmath}
\usepackage{multirow}
\usepackage{makecell}
\usepackage{setspace}
\usepackage{threeparttable}
\usepackage{color}
\usepackage{amsthm}
\usepackage{tabularx}
\usepackage{booktabs}
\newtheorem{definition}{Definition}

\def\BibTeX{{\rm B\kern-.05em{\sc i\kern-.025em b}\kern-.08em
    T\kern-.1667em\lower.7ex\hbox{E}\kern-.125emX}}
\usepackage{balance}
\begin{document}
\title{
AIDC Microgrid Vulnerability Assessment Under Computing-Power Coordinated Attacks
}
\vspace{-30pt}
\author{
Ze Yu,~\IEEEmembership{Student Member,~IEEE,}
Hongwei Zhen,~\IEEEmembership{Student Member,~IEEE,}
Chao Shen,~\IEEEmembership{Student Member,~IEEE,}
Mingyang Sun,~\IEEEmembership{Senior Member,~IEEE}\vspace{-30pt}
}

\markboth{Submitted to xxx}
{Yu \MakeLowercase{\textit{et al.}}: AIDC Microgrid Vulnerability Assessment Under Computing-Power Coordinated Attacks}

\maketitle

\begin{spacing}{0.85}
\begin{abstract}
The rapid growth of large language model (LLM) services is expanding AI data centers (AIDCs), increasing electricity demand and associated carbon emissions. Renewable energy integration can mitigate these impacts but also strengthens the coupling between AIDC loads and inverter-interfaced generation, creating cross-domain cyber-physical vulnerabilities. Specifically, adversarial AI requests alter AIDC power demand, whereas inverter control tampering modifies source-side dynamics, and their combined impact on system stability varies with generation forecast and demand response uncertainties.
To this end, we propose an uncertainty-aware AIDC microgrid vulnerability assessment framework under computing-power coordinated attacks. First, the framework maps adversarial AI requests to AIDC power variations and represents uncertainties in attack-induced demand responses and photovoltaic (PV) forecasts through confidence-weighted realizations. Then, impedance-based stability analysis combines these realizations with bounded inverter parameter tampering to construct attack reachable domains and identify critical attack time windows. Furthermore, a separate criterion identifies fixed coordinated attack vectors that retain destabilizing capability throughout each selected window. Case studies demonstrate that, unlike either attack component applied alone, coordinated attacks within identified critical windows induce sustained inverter frequency oscillations with peak absolute deviations exceeding 20\% of nominal frequency, whereas the evaluated out-of-window response remains bounded. The proposed method further identifies critical attack windows and the associated coordinated attack vectors. 
\end{abstract}


\vspace{-15pt}
\section{Introduction}

\IEEEPARstart{T}HE rapid growth of large language model (LLM) services is accelerating the expansion of AI data centers (AIDCs), which provide the computing infrastructure for large-scale AI training and inference. 
This expansion places increasing demands on power systems because AIDCs consume large and rapidly increasing amounts of electricity, raising concerns regarding associated carbon emissions. 
In 2025, electricity consumption by AIDCs increased by 50\%, compared with 17\% growth in total data center demand, which reached approximately 485 TWh. The IEA projects total demand to reach approximately 950 TWh by 2030, equivalent to around 3\% of global electricity consumption, and electricity consumption by AI-focused facilities to triple over the same period \cite{iea2026keyquestions}.
In the United States, EPRI projects that data centers could account for 9\%-17\% of national electricity consumption by 2030, up from 4\%-5\% \cite{epri2026powering}.

To reduce the emissions associated with this growth, AIDCs are increasingly being coupled with renewable generation and energy management systems. This transition is supported by policy targets, corporate decarbonization commitments, and power system planning that promote a higher share of clean electricity in AIDC operation \cite{ndrc2023_eastdata_westcompute_en,eu2024_datacentre_rating}. However, increased reliance on renewable generation and power electronic interfaces increases the sensitivity of AIDC operation to grid constraints and converter dynamics. Consequently, the stability of an AIDC microgrid depends jointly on the AIDC load condition, renewable generation operating point, and inverter control characteristics. 


Nevertheless, AIDC microgrids with inverter-interfaced renewable generation introduce cyberattack surfaces in both the computing infrastructure and the control channels associated with the IBRs \cite{mohsenian2011distributed,dafarl}. AIDC operation depends on interconnected computing and communication infrastructure, whereas IBRs regulate voltage, frequency, and power exchange through measurement and control signals. Attacks on computational infrastructure can disrupt workload execution and resource management, while the manipulation of IBR measurements or control signals can impair voltage and frequency regulation. Hence, characterizing how such attacks affect AIDC workloads and inverter control is essential for evaluating the security and stability of AIDC microgrids.




Recent studies indicate that AIDC infrastructures are exposed to diverse cyberattack surfaces. At the hardware level, Nazaraliyev \textit{et al.} \cite{nazaraliyev2025not} demonstrated side channel and denial-of-service attacks against shared GPUs, with application slowdown exceeding 4.8 times. At the cloud resource level, Manzoor \textit{et al.} \cite{manzoor2024enabling} developed ThreatPro to analyze dynamic interconnections and multi-layer attack propagation throughout virtual machine lifecycles. For network monitoring, Yuan \textit{et al.} \cite{yuan2024hybrid} proposed a spatiotemporal framework for detecting cyberattack traffic in cloud data center networks. The orchestration layer introduces additional risks, as Wang \textit{et al.} \cite{wang2025losing} revealed that insufficient access control and rate limiting in Kubernetes control plane interfaces can cause sensitive data leakage and excessive CPU consumption. To address threats involving both computing and forwarding components, Itani \textit{et al.} \cite{itani2025dcguard} developed a holistic approach for detecting and isolating malicious virtual machines and network switches in multi-tenant data centers. At the AI service layer, Zhang \textit{et al.} \cite{zhang2025crabs} showed that black-box LLM denial-of-service prompts can increase response latency by more than 250 times while exhausting GPU and memory resources.

From the power system perspective, there has also been significant research focusing on the cybersecurity and resilient control of microgrids. Tabassum \textit{et al.} \cite{tabassum2024cyber} developed an autoencoder-based method for detecting cyber-physical anomalies in microgrids. To mitigate the attack impact, Liu \textit{et al.} \cite{liu2024detection} proposed a recursive scheme to mitigate the effects of secondary false data injection attacks. For islanded AC microgrids, Vaishnav \textit{et al.} \cite{vaishnav2024auxiliary} further extended this research to distributed resilient control by combining the detection of bounded stealthy Byzantine attacks with frequency restoration and proportional active power sharing. Considering confidentiality in addition to attack resilience, Wang \textit{et al.} \cite{wang2025attack} integrated attack resilience and privacy preservation into distributed secondary control. Beyond communication and data integrity attacks, Kontou \textit{et al.} \cite{kontou2025exploiting} investigated the control interaction between grid-following and grid-forming inverters and showed that PLL attacks can induce severe voltage sags and instability, which can be mitigated through adaptive droop tuning.

At the IBR level, extensive research has further investigated cyberattacks against IBRs and their effects on power system stability. Musleh \textit{et al.} \cite{musleh2024experimental} experimentally evaluated security and stability vulnerabilities in commercial solar inverters. At the local control level, Wang and Pal \cite{wang2023destabilizing} used adversarial deep reinforcement learning to identify destabilizing attacks on droop control gains and derive corresponding robust defense strategies. Considering more general parameter tampering scenarios, Yu \textit{et al.} \cite{dafarl} developed a fuzzing-assisted reinforcement learning method to identify critical inverter parameters and construct destabilizing attacks. At the dispatch interface, Zhen \textit{et al.} \cite{zhen2026admittance} proposed an admittance-guided method to identify dispatch commands that induce severe subsynchronous oscillations while remaining within nominal operating limits. Focusing on the synchronization loop, Bamigbade \textit{et al.} \cite{bamigbade2023cyberattack} showed that PLL manipulation can cause voltage violations and system-wide power angle instability. At the broader system control level, Wang \textit{et al.} \cite{wang2023design} introduced modal resonance-oriented attacks that inject malicious signals into wide-area damping controllers to excite open-loop oscillatory modes.   


Despite the growing coupling between AIDCs and power systems, existing research has focused primarily on the flexibility of AIDC loads, while the cybersecurity implications arising from this coupling remain overlooked. At the power supply level, He \textit{et al.} \cite{he2024analysis} analyzed hybrid renewable power supply configurations for data centers and optimized the tradeoff between renewable penetration and electricity cost. At the dynamic modeling level, Gyang \textit{et al.} \cite{gyang2025dynamic} developed and validated a data center model incorporating electrical and thermal dynamics for evaluating grid disturbances, fault ride-through, and demand response. For transient stability analysis, Jimenez-Ruiz \textit{et al.} \cite{jimenez2025data} modeled the effects of UPS dynamics, cooling loads, and pulsing AI workloads on power system behavior. At the grid service level, Colangelo \textit{et al.} \cite{colangelo2026ai} demonstrated that coordinated AI workloads can provide responsive flexibility, while Ren \textit{et al.} \cite{ren2026grid} quantified the potential of flexible data center loads to support grid frequency stability.

Taken together, existing studies have separately investigated cybersecurity threats in AIDC infrastructures, microgrid cybersecurity, the stability impacts of cyberattacks on IBRs, and the flexibility potential of AIDC loads in power systems. Consequently, cross-domain cyber-physical vulnerabilities in AIDC microgrids remain insufficiently characterized. Specifically, adversarial AI requests can increase computational demand and thereby alter AIDC power demand without direct access to electrical equipment \cite{shumailov2021sponge,zhang2026resource}. Concurrently, an adversary may tamper with inverter control parameters, thereby modifying source-side impedance characteristics and the stability margin. It therefore remains unclear whether bounded attacks coordinated across these domains can create destabilizing source-load impedance interactions, particularly when neither attack component is individually destabilizing.

Moreover, this cross-domain problem is further complicated by uncertainty in both renewable generation and attack-induced AIDC demand responses. Renewable resource availability and ambient conditions determine the inverter operating point. Therefore, forecast errors render the future source impedance uncertain. Adversarial AI requests designed to increase resource consumption may also produce uncertain AIDC demand responses. The resulting power variation depends on prompt characteristics, model serving conditions, task allocation, and cooling dynamics. These uncertainties complicate the identification of feasible attack times and vectors whose destabilizing effects persist across plausible operating conditions.


To address these gaps, this paper develops an uncertainty-aware framework for assessing AIDC microgrid vulnerability under computing-power coordinated attacks. Specifically, the framework captures the joint cyber-physical effects of AI-induced demand manipulation and inverter control parameter attacks. The framework addresses two key challenges: 1) modeling the coupled cyber-physical risk arising from the interaction between AIDC demand and power system dynamics, which requires a joint representation of the two domains, and 2) assessing system vulnerability over future time horizons under joint uncertainty in operating conditions and AIDC demand responses induced by attacks.

The main contributions of this paper can be summarized as follows.
\begin{itemize}    
    \item To the best of our knowledge, this is the first study to formulate a cross-domain cyber-physical coordinated attack model for AIDC microgrids by integrating AI-induced demand manipulation with bounded PV inverter control parameter tampering. The model maps adversarial AI requests to workload-driven variations in AIDC power demand and equivalent impedance, thereby capturing their coupled effect on small-signal stability.

    \item To address uncertainties in renewable generation and AI-induced demand responses, an uncertainty-aware vulnerability assessment framework for AIDC microgrid is proposed. For each joint realization, the framework constructs attack reachable domains of critical closed-loop eigenvalues and defines an attackability score that quantifies the confidence-weighted prevalence of attackable conditions.
    
    \item A two-stage procedure is developed to identify attack time windows and their corresponding coordinated attack vectors. The attackability score determines when the system is vulnerable, whereas a window-level criterion selects vectors that satisfy a prescribed confidence threshold throughout each selected interval. This window–vector characterization distinguishes persistent vulnerabilities from isolated worst-case realizations.

    
\end{itemize}

\begin{figure}[h]
\centering
\includegraphics[width=0.4\textwidth]{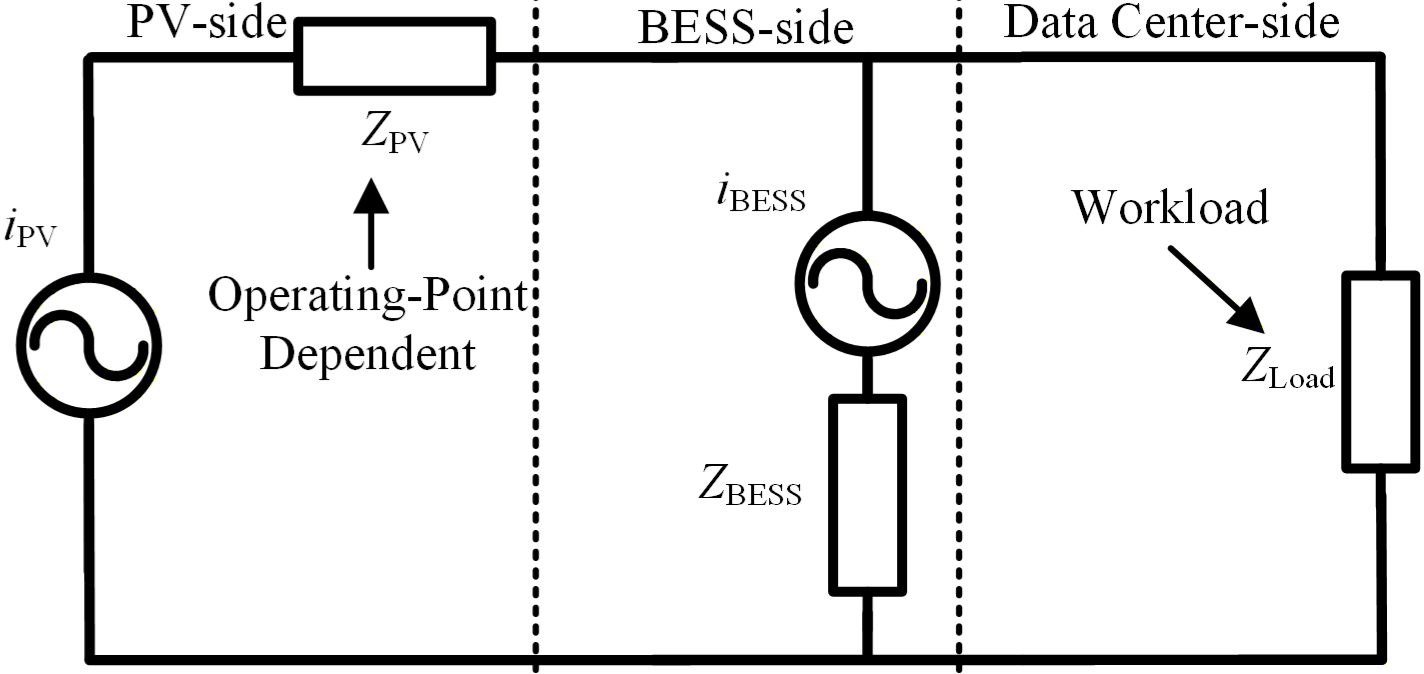}
\caption{Architecture of the studied AIDC microgrid.}
\label{system_model}
\end{figure}

\vspace{-20pt}
\section{System Model}
This section introduces the system model considered in this paper, which consists of three main components: the photovoltaic (PV) model, the AIDC load model, and the cyber-physical threat model. The overall architecture of the studied system is illustrated in Fig.~\ref{system_model}.

\vspace{-10pt}
\subsection{PV Model}
\label{subsec:pv_model}

The battery energy storage system (BESS) buffers the power exchanged by the PV subsystem and AIDC load, permitting their partial decoupling over the time scale considered. Specifically, the BESS primarily absorbs short-term AIDC workload variations, limiting their direct effect on the PV operating point. The PV model is therefore parameterized by irradiance and the inverter operating point.


\subsubsection{Steady-State PV Power Model}

Let $G$ denotes the solar irradiance, the steady-state PV output power can be expressed as a function of irradiance:
\begin{equation}
P_{\mathrm{pv},0}=P_{\mathrm{pv}}(G_0)
\label{eq:pv_power_steady}
\end{equation}
where $G_0$ denotes the nominal irradiance and $P_{\mathrm{pv},0}$ denotes the corresponding steady-state PV power. By linearizing \eqref{eq:pv_power_steady} around $G_0$, the incremental PV power variation is given by
\begin{equation}
\Delta P_{\mathrm{pv}}
=
\left.\frac{\partial P_{\mathrm{pv}}}{\partial G}\right|_{G_0} \cdot \Delta G
\label{eq:pv_power_sensitivity}
\end{equation}

Let $V_{\mathrm{pv},0}$ and $I_{\mathrm{pv},0}$ denote the steady-state PV terminal voltage and current, respectively, such that
\begin{equation}
P_{\mathrm{pv},0}=V_{\mathrm{pv},0} I_{\mathrm{pv},0}
\label{eq:pv_operating_point}
\end{equation}
Accordingly, the operating point perturbations induced by an irradiance variation are given by
\begin{equation}
\Delta V_{\mathrm{pv},0}=
\left.\frac{\partial V_{\mathrm{pv}}}{\partial G}\right|_{G_0} \cdot \Delta G
\label{eq:pv_operating_point_sensitivity_V}
\end{equation}
\begin{equation}
\Delta I_{\mathrm{pv},0}=
\left.\frac{\partial I_{\mathrm{pv}}}{\partial G}\right|_{G_0} \cdot \Delta G
\label{eq:pv_operating_point_sensitivity_I}
\end{equation}

Equations \eqref{eq:pv_power_sensitivity}--\eqref{eq:pv_operating_point_sensitivity_I} show that an irradiance variation changes both the available PV power and the PV unit operating point. The resulting operating point shift causes the associated impedance variation \cite{van2025analyzing}.

\subsubsection{Small-Signal Linearization and PV Nodal Impedance}

To characterize the dynamic behavior of the PV node, the PV array and its power electronic interface are modeled as a parameterized nonlinear system. Let $x_{\mathrm{pv}}$ denotes the state vector, $u_{\mathrm{pv}}$ denotes the nodal port voltage input, and $i_{\mathrm{pv}}$ denotes the injected nodal current. The PV subsystem can then be described as
\begin{equation}
\dot{x}_{\mathrm{pv}}
=
f_{\mathrm{pv}}(x_{\mathrm{pv}},u_{\mathrm{pv}},G),
\qquad
i_{\mathrm{pv}}
=
h_{\mathrm{pv}}(x_{\mathrm{pv}},u_{\mathrm{pv}},G)
\label{eq:pv_nonlinear_model}
\end{equation}
where $f_{\mathrm{pv}}(\cdot)$ describes the nonlinear internal dynamics of the PV subsystem, and $h_{\mathrm{pv}}(\cdot)$ maps its states and inputs to the PV node output current.

For a given irradiance level $G_0$, let $\left(x_{\mathrm{pv},0},u_{\mathrm{pv},0}\right)$ denote the corresponding equilibrium point satisfying
\begin{equation}
f_{\mathrm{pv}}(x_{\mathrm{pv},0},u_{\mathrm{pv},0},G_0)=0
\label{eq:pv_equilibrium}
\end{equation}

Linearization of \eqref{eq:pv_nonlinear_model} about $\left(x_{\mathrm{pv},0},u_{\mathrm{pv},0}\right)$ yields the small-signal model
\begin{equation}
\Delta \dot{x}_{\mathrm{pv}}
=
A_{\mathrm{pv}}(G_0)\Delta x_{\mathrm{pv}}
+
B_{\mathrm{pv}}(G_0)\Delta u_{\mathrm{pv}}
\label{eq:pv_ss_state}
\end{equation}
\begin{equation}
\Delta i_{\mathrm{pv}}
=
C_{\mathrm{pv}}(G_0)\Delta x_{\mathrm{pv}}
+
D_{\mathrm{pv}}(G_0)\Delta u_{\mathrm{pv}}
\label{eq:pv_ss_output}
\end{equation}
where $A_{\mathrm{pv}}$, $B_{\mathrm{pv}}$, $C_{\mathrm{pv}}$, and $D_{\mathrm{pv}}$ are the linearized PV-subsystem matrices parameterized by the operating point $G_0$.

Applying the Laplace transform to \eqref{eq:pv_ss_state}--\eqref{eq:pv_ss_output} yields the nodal small-signal admittance
\begin{equation}
Y_{\mathrm{pv}}(s,G_0)
\!=\!
C_{\mathrm{pv}}(G_0)
\big(sI\!-\!A_{\mathrm{pv}}(G_0)\big)^{-1}\!
B_{\mathrm{pv}}(G_0)
\!+\!
D_{\mathrm{pv}}(G_0)
\label{eq:pv_admittance}
\end{equation}
The corresponding nodal small-signal impedance is
\begin{equation}
Z_{\mathrm{pv}}(s,G_0)
=
Y_{\mathrm{pv}}^{-1}(s,G_0)
\label{eq:pv_impedance}
\end{equation}

\subsection{AIDC Model}

\subsubsection{Aggregate Power Demand Model}

The AIDC node is modeled as a workload-dependent load. Its information technology (IT) power consumption is represented by
\begin{equation}
P_{\mathrm{IT}}(t)
=
P_{\mathrm{idle}}
+
\kappa C(t)
\label{eq:dc_imp_it_power}
\end{equation}
where $P_{\mathrm{idle}}$ denotes the idle IT power, $\kappa>0$ denotes the computation-to-power coefficient, and $C(t)$ denotes the aggregate computing demand, which is mapped to the corresponding active and reactive power demands of the AIDC.

Assuming that cooling power varies proportionally with computing demand, the aggregate AIDC power demand is represented as
\begin{equation}
P_{\mathrm{dc}}(t)
=
\alpha_0
+
\alpha_1 C(t)
\label{eq:dc_imp_affine_power}
\end{equation}
where $\alpha_0$ denotes the baseline power demand and $\alpha_1$ denotes the effective workload-to-power conversion coefficient.

\subsubsection{PCC-Equivalent Impedance Model}

For system-level stability analysis, the apparent power consumed by the AIDC at the point of common coupling (PCC) is expressed as
\begin{equation}
S_{\mathrm{dc}}(t)
=
P_{\mathrm{dc}}(t)+jQ_{\mathrm{dc}}(t)
\label{eq:dc_imp_component_power}
\end{equation}
where $P_{\mathrm{dc}}(t)$ and $Q_{\mathrm{dc}}(t)$ denote the aggregate active and reactive power demands of the AIDC at time $t$, respectively.

Then, the AIDC load at the PCC is modeled by an equivalent series $R$-$L$ branch \cite{gyang2025dynamic}. The corresponding resistance and inductance are
\begin{equation}
R_{\mathrm{dc}}(t)
=
\frac{P_{\mathrm{dc}}(t)V_{\mathrm{rb}}^2}
{P_{\mathrm{dc}}^2(t)+Q_{\mathrm{dc}}^2(t)}
\label{eq:dc_imp_component_resistance}
\end{equation}
\begin{equation}
L_{\mathrm{dc}}(t)
=
\frac{Q_{\mathrm{dc}}(t)V_{\mathrm{rb}}^2}
{\omega\left(P_{\mathrm{dc}}^2(t)+Q_{\mathrm{dc}}^2(t)\right)}
\label{eq:dc_imp_component_inductance}
\end{equation}
where $V_{\mathrm{rb}}$ denotes the rated bus voltage, and $\omega$ denotes the nominal grid angular frequency.

The corresponding equivalent impedance is therefore
\begin{equation}
Z_{\mathrm{dc}}(s,t)=R_{\mathrm{dc}}(t)+sL_{\mathrm{dc}}(t)
\label{eq:dc_imp_component_impedance}
\end{equation}

This baseline model provides a physically interpretable impedance determined by the aggregate active and reactive power demands, but does not capture all frequency-dependent characteristics. To capture these effects in case studies, the AIDC impedance is further represented by a fitted model constructed from a real-world dataset\cite{garridozafra2025datacenterharmonicdataset}. Specifically, the impedance is expressed as
\begin{equation}
    Z_{\mathrm{dc}}(s,t) = Z_{\mathrm{fit}}(s,P_{\mathrm{dc}}(t),Q_{\mathrm{dc}}(t))
\end{equation}
where $Z_{\mathrm{fit}}(\cdot)$ denotes the fitted impedance model that maps the frequency variable $s$ and the AIDC load condition to the corresponding AIDC impedance.

\begin{table}[t]
\footnotesize 
\centering 
\caption{Threat Model Summary} 
\label{tab:threat_model} 
\renewcommand{\arraystretch}{1.12} 
\begin{tabular}
{p{0.25\linewidth} p{0.65\linewidth}} 
\hline 
\textbf{Category} & \textbf{Assumptions} \\ 
\hline \textbf{Capability} & 1) Bounded tampering with inverter control parameters. 2) AI-induced demand manipulation.\\ 
\textbf{Knowledge} & 1) Historical operating measurements. 2) An impedance model estimated from measurements.  \\  
\textbf{Access} & 1) Compromised PV inverter parameter update interface. 2) User-level access to the AIDC service. 3) Read access to PMU. \\ 
\textbf{Constraints} & 1) Parameter perturbations remain within admissible bounds. 2) Demand manipulation attacks produce uncertain demand responses. 3) No physical access to the electrical infrastructure. \\ 
\hline 
\end{tabular} 
\end{table}

\vspace{-10pt}
\subsection{Threat Model and Attack Assumptions}
\label{subsec:threat_model}

\begin{figure*}[h]
\centering
\includegraphics[width=0.9\textwidth]{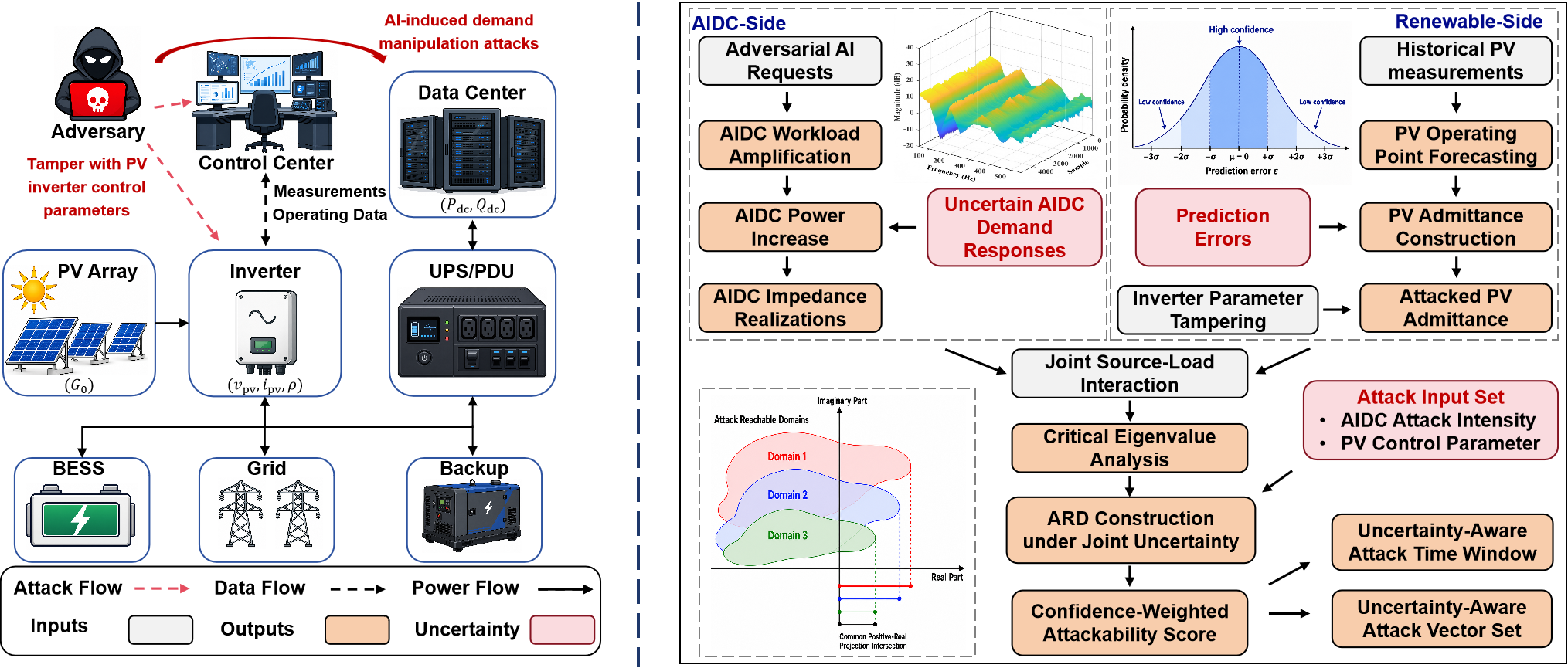}
\caption{Uncertainty-aware AIDC microgrid vulnerability assessment framework under computing-power coordinated attacks.}
\label{framework_overview}
\end{figure*}

As illustrated in Fig.~\ref{framework_overview}, a remote computing-power coordinated attack against an AIDC microgrid is considered. The attack seeks to reduce the small-signal stability margin by jointly modifying the PV inverter dynamics and AIDC load condition. The attacker is assumed to exploit two cyberattack surfaces: the communication-enabled PV inverter control interface and AI service interface of the AIDC. Direct physical access to electrical infrastructure is excluded. The threat model is summarized in Table~\ref{tab:threat_model}.

The attacker need not know the complete nonlinear system model but can estimate an impedance model from historical measurements \cite{zhen2026admittance}. Compromised access to PV inverter communication and parameter update interfaces permits modification of selected control parameters and acquisition of phasor measurement unit (PMU) data. Voltage and current phasors, together with historical operating trajectories, are used to forecast PV operating points and AIDC load conditions. These measurements also support inference of the impedance characteristics required for attack planning. The attacker also has user-level access to the AI service and can submit adversarial prompts that increase resource consumption. However, future PV forecast errors and demand response realizations remain unknown.

Within these access constraints, the attacker modifies selected inverter control parameters within admissible bounds \cite{dafarl}, thereby avoiding immediate protective action. The attacker cannot alter the system topology, override hardware or protection constraints, or manipulate environmental conditions or the energy storage state. Adversarial AI-service requests increase computational demand without directly controlling electrical equipment \cite{shumailov2021sponge}. Under the PCC equivalent load model, the resulting AIDC power variation modifies the load impedance observed by the microgrid. Because this response depends on workload scheduling, queuing, and cooling dynamics, the impedance variation is modeled as bounded and uncertain.

Although temporally coordinated, the inverter and load-side attacks are individually bounded and need not be destabilizing. Inverter control modifications may be ineffective when the stability margin is large, whereas adversarial AI requests alone cannot deterministically induce instability. However, their combination can reshape the coupling between source and load impedances and erode the margin, allowing bounded inverter perturbations to move the dominant closed-loop eigenvalue towards the stability boundary.

\vspace{-10pt}
\section{Uncertainty-Aware AIDC Microgrid Vulnerability Assessment Framework}
\subsection{Overview}

In this paper, we develop an uncertainty-aware AIDC microgrid vulnerability assessment framework under computing-power coordinated attacks. Fig.~\ref{framework_overview} illustrates the framework, which accounts for two uncertainty sources. The first is forecast-driven PV uncertainty, which affects the renewable generation operating point and the corresponding PV impedance trajectory. The second is attack-response uncertainty on the AIDC side, where adversarial AI requests induce workload-dependent demand variations with uncertain electrical responses. The framework represents future PV operating points using confidence-weighted realizations, models AI-induced demand manipulation as uncertain impedance changes, and evaluates their joint effect on the source-load impedance interaction. Based on this interaction, the framework performs closed-loop eigenvalue analysis, constructs attack reachable domains (ARDs)~\cite{zhen2026ard} under joint uncertainty, and identifies uncertainty-aware attack time windows and corresponding attack vectors.

\vspace{-10pt}
\subsection{AIDC Workload Formation and Attack-Induced Load Dynamics}

This subsection formulates the mapping from AI-service requests to AIDC power demand and characterizes the modification of this demand by adversarial signals. Training workloads are generally long-running and constrained by scheduling and accelerator locality \cite{jeon2019analysis}. By contrast, online inference exhibits bursty request arrivals, imposes strict latency requirements, and comprises distinct prefill and decode phases \cite{wang2025burstgpt,agrawal2024taming}. Let $\mathcal{R}_t$ denotes the set of AI-service requests arriving during observation interval $\Delta t$ at time $t$, the offered compute rate is defined as
\begin{equation}
    C^{\mathrm{off}}(t)
    =
    \frac{1}{\Delta t}\sum_{i\in\mathcal{R}_t}w_i
    \label{eq:aidc_offered_workload}
\end{equation}
where $w_i$ denotes the accelerator work associated with request $i$. Given queued workload $B_t$ and $n_t$ active accelerator groups, the realized compute rate is
\begin{equation}
    C(t)
    =
    \min\left\{
    C^{\mathrm{off}}(t)+\frac{B_t}{\Delta t},
    \overline{C}n_t
    \right\}
    \label{eq:aidc_realized_power}
\end{equation}

This formulation distinguishes the workload requested by users from that physically executed by the AIDC. Consequently, an abrupt increase in request arrivals may produce an immediate power ramp when computational headroom is available or a prolonged disturbance caused by queuing and delayed autoscaling under capacity saturation.

In addition to increasing request volume, an attacker may increase per-request computational workload by inducing long contexts, forced long outputs, end-of-sequence (EOS) suppression, repeated retrieval operations, and repeated tool calls \cite{shumailov2021sponge,liu2026inference,dong2025engorgio}. Let $\mathcal{T}_a=[t_a,t_a+T_a]$ denote the attack interval. The cumulative offered workload over this interval is defined as
\begin{equation}
    W_{\mathcal{T}_a}
    =
    \int_{\mathcal{T}_a}
    C^{\mathrm{off}}(t)\,\mathrm{d}t
    \label{eq:attack_interval_work}
\end{equation}

Let $W_{\mathcal{T}_a}^{0}$ and $W_{\mathcal{T}_a}^{\mathrm{atk}}$ denote the cumulative offered workloads under baseline and attack conditions, respectively. The resulting workload-amplification factor is
\begin{equation}
    \frac{
        W_{\mathcal{T}_a}^{\mathrm{atk}}
    }{
        W_{\mathcal{T}_a}^{0}
    }
    =
    1+r(A_w-1)+\rho_N A_b.
    \label{eq:interval_work_amplification}
\end{equation}
where $A_w$ is the per-request workload amplification factor, $A_b$ is the mean workload of an injected prompt normalized by that of a baseline request, $\rho_N$ is the ratio of injected to baseline requests, and $r$ is the fraction of baseline requests subject to complexity amplification. This expression quantifies submitted computational work rather than instantaneous executed workload.

The realized compute rate $C(t)$ determines the physical AIDC power response. When sufficient accelerator headroom is available, the additional work can be executed immediately. Otherwise, request queuing or rate limiting may defer execution until autoscaling provisions additional capacity. The resulting additional energy consumption and average power increase over the response horizon are expressed as
\begin{align}
    \Delta E_{\mathcal{H}_a}
    &=
    \int_{\mathcal{H}_a}
    \left[
        P_{\mathrm{dc}}^{\mathrm{atk}}(t)
        -
        P_{\mathrm{dc}}^{0}(t)
    \right]\mathrm{d}t
    \label{eq:attack_additional_energy}\\
    \Delta\overline{P}_{\mathcal{H}_a}
    &=
    \frac{\Delta E_{\mathcal{H}_a}}{T_r}
    \label{eq:average_and_peak_power}
\end{align}
where $P_{\mathrm{dc}}^{\mathrm{atk}}(t)$ and $P_{\mathrm{dc}}^{0}(t)$ denote the instantaneous AIDC power demands under attack and baseline conditions, respectively. $\mathcal{H}_a=[t_a,t_a+T_r]$ is the complete response horizon, and $T_r$ is the duration of the response horizon.

Assuming that all injected work is eventually processed and that the baseline workload remains approximately constant, the average relative power increase is approximated as
\begin{equation}
    \delta(u_{\mathrm{dc}})
    =
    \frac{
        \Delta\overline{P}_{\mathcal{H}_a}
    }{
        \overline{P}_{\mathrm{dc}}^{0}
    }
    \approx
    \eta_C
    \frac{T_a}{T_r}
    \left[
        r(A_w-1)+\rho_NA_b
    \right]
    \label{eq:average_power_amplification}
\end{equation}
where $\eta_C$ denotes the workload-sensitive fraction of the baseline AIDC power, and $u_{\mathrm{dc}}$ denotes the intensity of the demand manipulation attack, with parameters $T_a$, $r$, $A_w$, $\rho_N$, and $A_b$. The components of $u_{\mathrm{dc}}$ have distinct control properties. The request-submission duration $T_a$ is specified by the attacker, whereas $r$, $A_w$, $\rho_N$, and $A_b$ characterize the effective workload response associated with the submitted request pattern. These response parameters are not assumed to be precisely controlled by the attacker.

An increase in attack intensity raises the interval-average AIDC power, and this effect is distinct from the corresponding increase in cumulative computational work. Table~\ref{tab:ai_service_attack_surfaces} summarizes representative attack vectors against AI services and their potential effects on AIDC computing demand and electrical load. Despite differences in their computational mechanisms, these attacks modify the magnitude, duration, and spatial distribution of the realized AIDC load.

\begin{table*}[!t]
    \centering
    \caption{Representative Attack Vectors Against AI Services and Their Potential Effects on AIDC Computing Demand and Electrical Load.}
    \label{tab:ai_service_attack_surfaces}
    \scriptsize
    \renewcommand{\arraystretch}{1.15}
    \setlength{\tabcolsep}{3.5pt}

    \begin{tabular}{
        p{3.90cm}
        p{5.30cm}
        p{1.80cm}
        p{5.70cm}}
        \toprule
        \textbf{Attack surface}
        &
        \textbf{Computational mechanism}
        &
        \textbf{Time scale}
        &
        \textbf{Potential electrical manifestation}
        \\
        \midrule

        \textbf{Adversarial request flooding}
        \cite{wang2025burstgpt,xu2014power}
        \newline \emph{Public API access}
        &
        Coordinated prompts increase arrival rate, concurrency,
        batch occupancy, and queue length.
        &
        Seconds--minutes
        &
        When computing capacity is available, the higher request arrival rate can trigger the activation of additional resources and thereby produce a rapid increase in aggregate power demand.
        \\

        \textbf{Complex input attack}
        \cite{shumailov2021sponge}
        \newline \emph{Public API access}
        &
        Long contexts increase prefill work and KV-cache allocation.
        Sponge inputs induce inefficient execution.
        &
        Seconds--minutes
        &
        Accelerator utilization duration and cumulative energy consumption increase, whereas hardware limits constrain the increase in peak power.
        \\

        \textbf{Output length attack}
        \cite{dong2025engorgio,liu2026inference}
        \newline \emph{Query access}
        &
        EOS suppression and repetitive decoding increase the
        generated token count.
        &
        Seconds--minutes
        &
        The decoding load remains elevated, while token amplification
        primarily increases cumulative energy consumption and interval average power.
        \\

        \textbf{KV-cache and scheduler exhaustion}
        \cite{wang2026rethinking}
        \newline \emph{API access}
        &
        Memory saturation triggers blocking, eviction, preemption,
        and recomputation.
        &
        Seconds--minutes
        &
        Memory traffic and load oscillations increase. Throughput
        collapse can extend the disturbance without increasing peak
        power.
        \\

        \textbf{RAG amplification}
        \cite{liu2026inference}
        \newline \emph{Query-induced retrieval amplification}
        &
        Malicious queries or poisoned documents may trigger broader retrieval, repeated reranking, longer retrieved contexts, or additional generation steps.
        &
        Seconds--minutes
        &
        Computational activity may increase across storage, CPU, network, and GPU components, potentially producing a multicomponent, temporally distributed power response.
        \\

        \textbf{Toolchain amplification}
        \cite{luo2026autonomy,zhou2026beyond}
        \newline \emph{Prompt, content, and tool access}
        &
        Repeated model calls, tool invocations, context growth, and
        retries expand the execution trajectory.
        &
        Minutes--hours
        &
        Prolonged multiservice execution may increase cumulative energy consumption and sustain elevated AIDC power demand.
        \\



        \textbf{Sponge model poisoning}
        \cite{cina2025energy}
        \newline \emph{Training-pipeline access}
        &
        Malicious updates increase activation density and reduce
        accelerator efficiency during future inference.
        &
        Persistent until retraining
        &
        A persistent increase in inference energy shifts the baseline AIDC
        load.
        \\

        \textbf{Scheduling signal manipulation}
        \cite{wang2015proactive}
        \newline \emph{Potential attack surface}
        &
        Falsified price, carbon, latency, and congestion signals alter
        job timing and site selection.
        &
        Minutes--hours
        &
        Temporal workload shifting produces demand valleys and rebound peaks, whereas spatial
        workload shifting redistributes power demand among AIDC sites.
        \\

        \bottomrule
    \end{tabular}
\end{table*}

\vspace{-10pt}
\subsection{Coordinated Load-Side and Inverter-Side Attack Model}
\label{subsec:coordinated_attack_model}

Based on the threat model and attack assumptions introduced in Section~II, this subsection formulates the coordinated attack under uncertainty by mapping inverter control tampering and AIDC demand responses to the impedance-based small-signal stability assessment framework.

\subsubsection{Load-side attack and uncertain demand response}
AIDC demand variation is modeled as an attack-dependent response rather than a purely forecast-derived quantity. The resulting variation modifies the equivalent AIDC impedance and can reduce the source-load stability margin. Concurrent inverter control modifications alter the source impedance, and the combined effects may move the system across the stability boundary.

For a given load-side attack parameter vector $u_{\mathrm{dc}}$ at time $t+\tau$, the $m$-th realization of the resulting AIDC apparent power is represented as
\begin{equation}
S_{\mathrm{dc}}^{(m)}
=
S_{\mathrm{dc}}^{0}
\left[
1+
\delta
\left(
u_{\mathrm{dc}}
\right)
+
\epsilon_{\mathrm{dc}}^{(m)}
\right]
\label{eq:uncertain_dc_power_response}
\end{equation}
where $S_{\mathrm{dc}}^{0}$ denotes the baseline AIDC
apparent power, $\epsilon_{\mathrm{dc}}^{(m)}$ denotes the deviation of the $m$-th realized response from its nominal value, and $m$ indexes the sampled response realizations. The attack response is parameterized by the active power variation. The corresponding apparent power is then used to update the AIDC impedance, and its relative increase is assumed to follow the active power increase.

Then, each response realization is assigned a confidence weight:
\begin{equation}
c_{\mathrm{dc}}^{(m)}(t\!+\!\tau)
\!=\!
\mathcal{C}_{\mathrm{dc}}
\left(
 \epsilon_{\mathrm{dc}}^{(m)}(t\!+\!\tau)
\right),
\quad
0\!\le\! c_{\mathrm{dc}}^{(m)}(t\!+\!\tau)\!\le\! 1
\label{eq:llm_load_confidence}
\end{equation}
where $c_{\mathrm{dc}}^{(m)}$ denotes the relative probability weight assigned to the $m$th demand response realization, and $\mathcal{C}_{\mathrm{dc}}(\cdot)$ denotes the probability-weight mapping derived from the load uncertainty model. The corresponding attacked AIDC impedance is
\begin{equation}
Z_{\mathrm{dc}}^{(m)}(s,t\!+\!\tau)
\!=\!
Z_{\mathrm{fit}}
\left(
s,
S_{\mathrm{dc}}^{(m)}(t\!+\!\tau,u_{\mathrm{dc}})
\right)
\label{eq:attacked_dc_impedance_uncertain}
\end{equation}

\subsubsection{PV-side uncertainty and inverter-side attack}
PV-side uncertainty arises from solar irradiance forecast errors that propagate to the predicted PV operating-point trajectory. Because a deterministic forecast cannot represent the range of plausible future conditions, multiple operating-point realizations are constructed, assigned confidence weights, and mapped to the corresponding impedance trajectories for vulnerability assessment.

The inverter-side attack is represented by modifications to the inverter control parameters. The corresponding attack vector is defined as
\begin{equation}
v_{\mathrm{pv}}^{\mathrm{atk}}(t+\tau)
=
\left\{
\Delta {\rho}_{1}(t+\tau),
\Delta {\rho}_{2}(t+\tau),
\ldots
\right\}
\label{eq:pv_attack_vector}
\end{equation}
where $\Delta\rho_j$ denotes the attack-induced modification applied to the $j$-th inverter control parameter.

For $v_{\mathrm{pv}}^{\mathrm{atk}}(t+\tau)$ and the $n$-th PV prediction-error realization, the attacked PV admittance is expressed as
\begin{equation}
\widetilde Y_{\mathrm{pv}}^{(n)}
(s,t\!+\!\tau,v_{\mathrm{pv}}^{\mathrm{atk}})
=
Y_{\mathrm{pv}}
\left(
s,
{\xi}_{\mathrm{pv}}^{(n)}(t\!+\!\tau|t),
v_{\mathrm{pv}}^{\mathrm{atk}}(t\!+\!\tau)
\right)
\label{eq:attacked_pv_admittance}
\end{equation}
where ${\xi}_{\mathrm{pv}}^{(n)}$ denotes the PV operating point under the $n$-th prediction-error realization. The realization is assigned a confidence weight $c_{\mathrm{pv}}^{(n)} \in [0,1]$.

The coordinated attack vector at time $t+\tau$ is defined as
\begin{equation}
a(t\!+\!\tau)
\!=\!
\left\{
v_{\mathrm{pv}}^{\mathrm{atk}}(t\!+\!\tau),
u_{\mathrm{dc}}(t\!+\!\tau)
\right\}
\in
\Omega_a(t\!+\!\tau)
\label{eq:coordinated_attack_vector_uncertain}
\end{equation}
where $\Omega_a(t+\tau)$ denotes the feasible coordinated attack set.

For a joint realization of PV prediction uncertainty and AI-induced demand response uncertainty, the attacked AIDC impedance and PV admittance yield the loop-gain matrix
\begin{equation}
L^{(n,m)}(s,t+\tau,a)
=
Z_{\mathrm{eq}}^{(m)}(s,t+\tau)
\widetilde Y_{\mathrm{pv}}^{(n)}(s,t+\tau,a)
\label{eq:attacked_loop_gain}
\end{equation}
where $Z_{\mathrm{eq}}^{(m)}$ denotes the equivalent external-microgrid impedance seen from the inverter under the $m$-th AI-induced demand response realization.

The closed-loop eigenvalues are defined as
\begin{equation}
\Lambda_{t+\tau}^{(n,m)}(a)\!=\!
\left\{
\lambda_i
\ \middle|\
\det
\left[
\mathbf{I}
\!+\!
L^{(n,m)}(\lambda_i,t+\tau,a)
\right]
\!=\!
0
\right\}
\label{eq:closed_loop_eigenvalue_set}
\end{equation}
where \(\mathbf{I}\) denotes the identity matrix.

\subsection{Uncertainty-Aware Attack Time Windows and Vectors}
\label{subsec:ua_attack_window_vector2}

Based on the coordinated attack model, this subsection identifies future intervals during which the system is vulnerable to coordinated attacks. For each uncertainty realization, the ARD characterizes whether a feasible coordinated attack can shift the critical closed-loop eigenvalue into the unstable region. Unlike deterministic attack window identification, the proposed formulation jointly accounts for PV operating point uncertainty and AI-induced demand response uncertainty.

For the $n$-th PV prediction-error realization and the $m$-th AI-induced demand response realization at future time $t+\tau$, the closed-loop eigenvalue with the largest real part is selected as
\begin{equation}
i^\star
\in
\arg\max_i
\operatorname{Re}
\left[
\lambda_i
\right],
\quad
\lambda_i\in
\Lambda_{t+\tau}^{(n,m)}(a).
\label{eq:critical_eigenvalue_index}
\end{equation}
The corresponding critical closed-loop eigenvalue is defined as
\begin{equation}
\lambda_{\mathrm{crit}}^{(n,m)}
(t+\tau,a)
=
\lambda_{i^\star}.
\label{eq:critical_closed_loop_eigenvalue}
\end{equation}

For the $n$-th PV prediction-error realization and the $m$-th demand response realization, the stability impact of a coordinated attack vector $a$ is quantified by the largest real part among the closed-loop eigenvalues:
\begin{equation}
\eta_{t+\tau}^{(n,m)}(a)
=
\operatorname{Re}
\left[
\lambda_{\mathrm{crit}}^{(n,m)}
(t+\tau,a)
\right]
\label{eq:ua_instability_margin}
\end{equation}
A positive value of $\eta_{t+\tau}^{(n,m)}(a)$ indicates that the coordinated attack vector $a$ places at least one closed-loop eigenvalue in the open right-half plane, thereby inducing small-signal instability.

Extending the analysis from a single attack vector to the feasible attack set yields the ARD for each uncertainty realization at time $t+\tau$:
\begin{equation}
\mathcal{D}_{t\!+\!\tau}^{(n,m)}
\!=\!
\left\{
\lambda_{\mathrm{crit}}^{(n,m)}
(t\!+\!\tau,a)
\! \middle| \!\
a \!\in\!\Omega_a(t\!+\!\tau)\!
\right\}
\!\subset\!\mathbb{C}
\label{eq:ua_attack_reachable_domain}
\end{equation}
This domain comprises all critical closed-loop eigenvalue locations reachable by feasible coordinated attack vectors under the $n$-th PV operating-point realization and the $m$-th AI-induced demand response realization.

Because small-signal stability is determined by the sign of the real part of the critical closed-loop eigenvalue, the ARD is projected onto the real axis:
\begin{equation}
\mathcal{P}_{t+\tau}^{(n,m)}
=
\left\{
\operatorname{Re}(z)
\mid
z\in
\mathcal{D}_{t+\tau}^{(n,m)}
\right\}
\subset\mathbb{R}
\label{eq:ua_real_projection}
\end{equation}
The binary attackability indicator is defined as
\begin{equation}
\chi_{t+\tau}^{(n,m)}
=
\mathbb{I}
\left[
\mathcal{P}_{t+\tau}^{(n,m)}
\cap
(0,\infty)
\neq
\emptyset
\right]
\label{eq:ua_attackability_indicator_projection}
\end{equation}
where $\mathbb{I}[\cdot]$ denotes the indicator function. Thus, $\chi_{t+\tau}^{(n,m)}=1$ indicates that, under the $n$-th PV operating-point realization and the $m$-th AI-induced demand response realization, at least one feasible coordinated attack vector can shift the critical closed-loop eigenvalue into the unstable region.

In the present implementation, the PV prediction error and AI-induced demand response are assumed to be conditionally independent given the scheduled BESS operating mode, yielding a tractable confidence-weighted formulation. Correlated uncertainties can instead be represented by replacing the product weight with a joint probability model. Under the assumption, the joint confidence weight is defined as
\begin{equation}
    \gamma_{t+\tau}^{(n,m)}=c_{\mathrm{pv}}^{(n)}(t+\tau)c_{\mathrm{dc}}^{(m)}(t+\tau)
\end{equation}

\begin{definition}[Uncertainty-Aware Attack Time Window]
Given a confidence threshold $\alpha\!\in\![0,1]$, a future time instant $t+\tau$ is classified as $\alpha$-confidence attackable if
\begin{equation}
A_{t+\tau}
\ge
\alpha
\label{eq:ua_attackable_time_condition_conf}
\end{equation}
where $A_{t+\tau}$ denotes the confidence-weighted attackability score at time $t+\tau$, defined as
\begin{equation}
A_{t+\tau}
=
\frac{
\sum_{n=1}^{N_{t+\tau}}
\sum_{m=1}^{M_{t+\tau}}
\gamma_{t+\tau}^{(n,m)}
\chi_{t+\tau}^{(n,m)}
}{
\sum_{n=1}^{N_{t+\tau}}
\sum_{m=1}^{M_{t+\tau}}
\gamma_{t+\tau}^{(n,m)}
}
\label{eq:ua_confidence_weighted_attackability}
\end{equation}
where $M_{t+\tau}$ and $N_{t+\tau}$ denote the numbers of sampled demand response and PV prediction-error realizations, respectively, at time $t+\tau$.

The corresponding $\alpha$-confidence attackable time set is
\begin{equation}
\mathcal{T}_{\alpha}^{\mathrm{UA}}
=
\left\{
t+\tau
\in
[t+1,t+H]
\ \middle|\
A_{t+\tau}
\ge
\alpha
\right\}
\label{eq:ua_attackable_time_set_conf}
\end{equation}
The consecutive intervals contained in $\mathcal{T}_{\alpha}^{\mathrm{UA}}$ are called $\alpha$-confidence uncertainty-aware attack time windows.
\end{definition}

The score $A_{t+\tau}\in[0,1]$ quantifies the normalized confidence weight of joint uncertainty realizations under which at least one feasible coordinated attack can destabilize the system at time $t+\tau$. A larger score therefore indicates that attackability persists across a greater confidence-weighted proportion of plausible future conditions. Accordingly, $A_{t+\tau}$ characterizes the prevalence, rather than the severity, of vulnerability. Because the effective attack vector may differ among realizations, the score does not represent the success probability of any specific attack vector.

Following identification of the uncertainty-aware attack time windows, the analysis determines whether a single coordinated attack vector remains effective throughout each window. For $a\in\Omega_a(t+\tau)$, its confidence-weighted effectiveness at time $t+\tau$ is defined as
\begin{equation}
S_{t+\tau}(a)
=
\frac{
\sum_{n=1}^{N_{t+\tau}}
\sum_{m=1}^{M_{t+\tau}}
\gamma_{t+\tau}^{(n,m)}
\mathbb{I}
\left[
\eta_{t+\tau}^{(n,m)}(a)>0
\right]
}{
\sum_{n=1}^{N_{t+\tau}}
\sum_{m=1}^{M_{t+\tau}}
\gamma_{t+\tau}^{(n,m)}
}
\label{eq:ua_attack_vector_score}
\end{equation}

\begin{definition}[Uncertainty-Aware Attack Vector]
For an uncertainty-aware attack time window $W=[t_s,t_e]$, a coordinated attack vector $a_W^{\mathrm{UA}}$ is defined as a $\beta$-confidence uncertainty-aware attack vector over $W$ if
\begin{equation}
a_W^{\mathrm{UA}}
\in
\bigcap_{t\in W}
\Omega_a(t)
\label{eq:ua_attack_vector_feasibility_conf}
\end{equation}
and
\begin{equation}
\min_{t\in W}
S_{t}
\left(
a_W^{\mathrm{UA}}
\right)
\ge
\beta
\label{eq:ua_attack_vector_condition_conf}
\end{equation}
where $\beta\in[0,1]$ is the prescribed attack vector confidence threshold. For $\beta=1$, the criterion requires the same attack vector to destabilize the system at every time instant in $W$.
\end{definition}

The set of all $\beta$-confidence uncertainty-aware attack vectors over $W$ is
\begin{equation}
\mathcal{A}_{W,\beta}^{\mathrm{UA}}
=
\left\{
a\in
\bigcap_{t\in W}
\Omega_a(t)
\ \middle|\
\min_{t\in W}
S_{t}(a)
\ge
\beta
\right\}
\label{eq:ua_attack_vector_set}
\end{equation}

Uncertainty-aware attack windows identify periods during which the attackability score of an AIDC microgrid exceeds a prescribed confidence threshold, whereas the corresponding vectors identify feasible coordinated attacks that remain effective under joint PV prediction and AI-induced demand response uncertainties. Relative to deterministic analysis, this characterization distinguishes persistent vulnerabilities from isolated worst-case conditions and provides a basis for prioritizing security detection and defensive measures during vulnerable operating periods.

\vspace{-10pt}
\section{Case Studies}
\subsection{Experiment Setup}

Case studies are conducted using an AIDC microgrid model developed in MATLAB/Simulink. The simulated system comprises a PV farm, a BESS, a grid interface, and an AIDC load. Consistent with Gaussian error models commonly used in probabilistic solar forecasting and power system uncertainty analysis \cite{yoo2024modeling}, the PV forecast error and demand response uncertainty were represented by zero-mean Gaussian random variables. The corresponding standard deviations were set to 15\% of the predicted PV output and 5\% of the nominal load variation, respectively. The PV generation profile was obtained from the public dataset in \cite{OEDI_Dataset_4568}. The nominal values of the inverter control parameters subject to attack are listed in Table~\ref{tab:setup}.

To characterize data center impedance variations across load conditions, the measured harmonic voltage and current phasors at the PCC from the university data-center dataset reported in \cite{garridozafra2025datacenterharmonicdataset} were used. The harmonic impedance at each sampling instant was computed as the ratio of the measured voltage phasor to the corresponding current phasor. An operating-point-dependent impedance model was then fitted to the resulting samples and incorporated into the proposed framework.

\begin{figure}[t]
    \centering
    \includegraphics[width=0.45\textwidth]{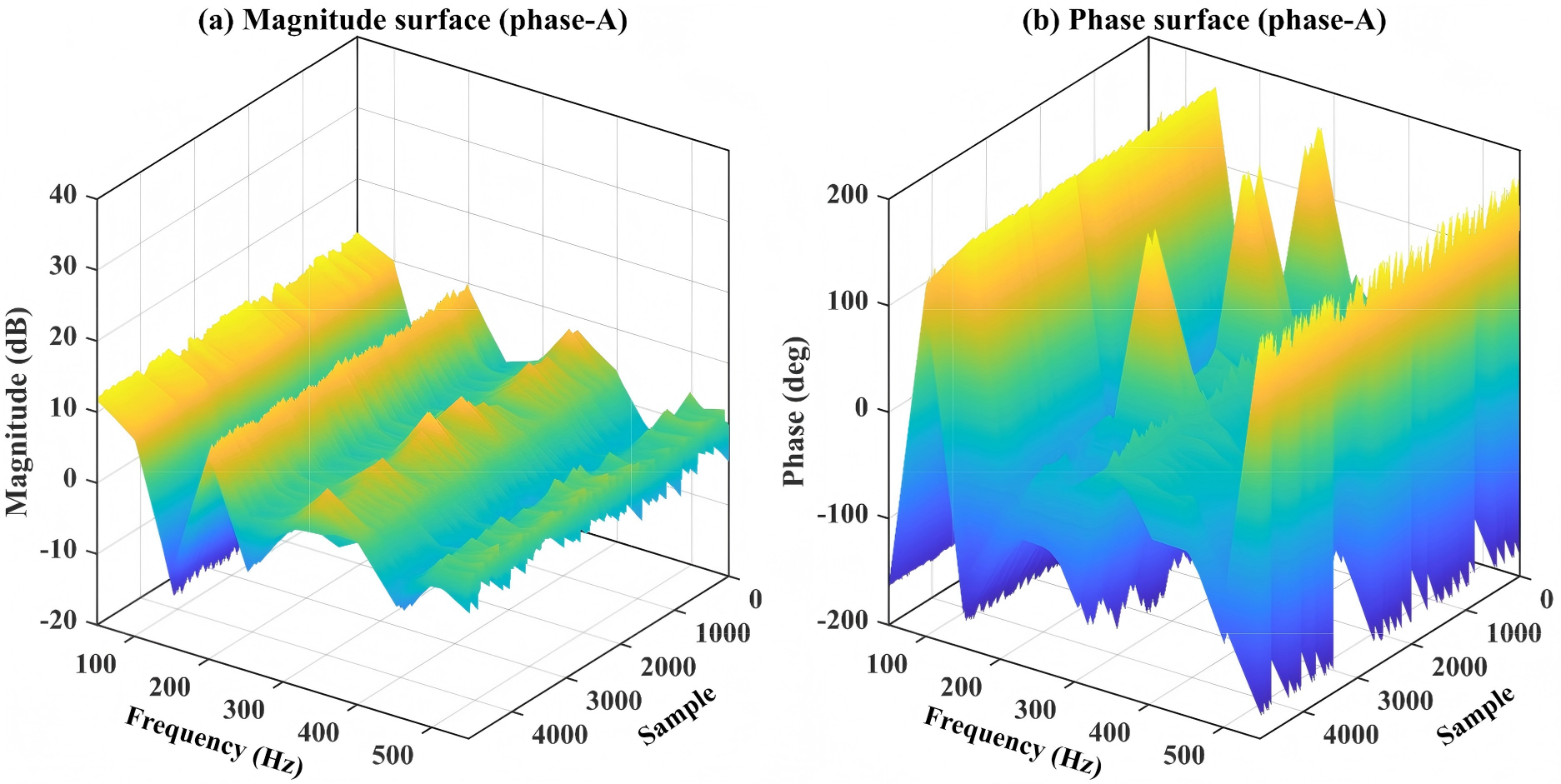}
    \caption{Measured harmonic impedance characteristics of a university data center under varying load conditions.}
    \label{fig:dc_impedance_bode}
\end{figure}

Fig.~\ref{fig:dc_impedance_bode}(a) shows variation in the measured impedance magnitude at several harmonic frequencies. At 50~Hz, the impedance magnitude had a median of 12.388~dB and a standard deviation of 0.723~dB. Over the same 5,000 samples, the largest changes between adjacent samples reached 69\% in impedance magnitude at 250~Hz and 119.7$^\circ$ in phase at 300~Hz. The results indicate that the data center load characteristics vary across the sampled conditions. Fig.~\ref{fig:dc_impedance_bode}(b) shows a similar phenomenon in the phase response. The phase surface exhibits large variations and discontinuities across different frequency ranges, suggesting that a fixed-impedance representation may not capture the observed variation. The measured harmonic impedance is not directly treated as the small-signal impedance. Instead, it is used to construct an operating-point-dependent impedance surrogate, which serves as the small-signal approximation in the stability analysis.


\begin{table}[!t]
\centering
\footnotesize
\caption{Nominal inverter control parameters subject to attack.}
\label{tab:setup}
\begin{tabular}{c c c}
\hline

\textbf{Parameter} & \textbf{Description} & \textbf{Value} 
\\
\hline
$k_{pv}$ & Proportional gain of the voltage-control loop & 0.433
\\

$k_{iv}$& Integral gain of the voltage-control loop&203.98
\\

$k_{pi}$& Proportional gain of the current-control loop& 0.525
\\
$J$& Virtual inertia coefficient of the inverter control & 0.116
\\
$D_p$& Active power damping coefficient& 36192
\\
\hline
\end{tabular}
\end{table}

\begin{table}[!t]
    \centering
    \caption{Illustrative effects of attack parameters on AIDC workload and interval-averaged
    facility power.}
    \label{tab:aidc_power_impact}
    \small
    \setlength{\tabcolsep}{5pt}
    \begin{tabular}{cccccccc}
        \toprule
        $A_w$
        & $r$
        & $\rho_N$
        & $A_b$
        & $\eta_C$
        & $T_a/T_r$
        & Workload
        & Power \\
        \midrule
        13.12 & 4\% & 5\%  & 1.00 & 0.40 & 1.00
        & 53.5\%  & 21.4\% \\

        13.12 & 6\% & 15\% & 1.10 & 0.40 & 0.90
        & 89.2\%  & 32.1\% \\

        13.12 & 7\% & 30\% & 1.30 & 0.40 & 0.82
        & 123.8\% & 40.6\% \\

        13.12 & 9\% & 45\% & 1.50 & 0.40 & 0.75
        & 176.6\% & 53.0\% \\
        \bottomrule
    \end{tabular}
\end{table}

\begin{figure}[t]
    \centering
    \includegraphics[width=0.45\textwidth]{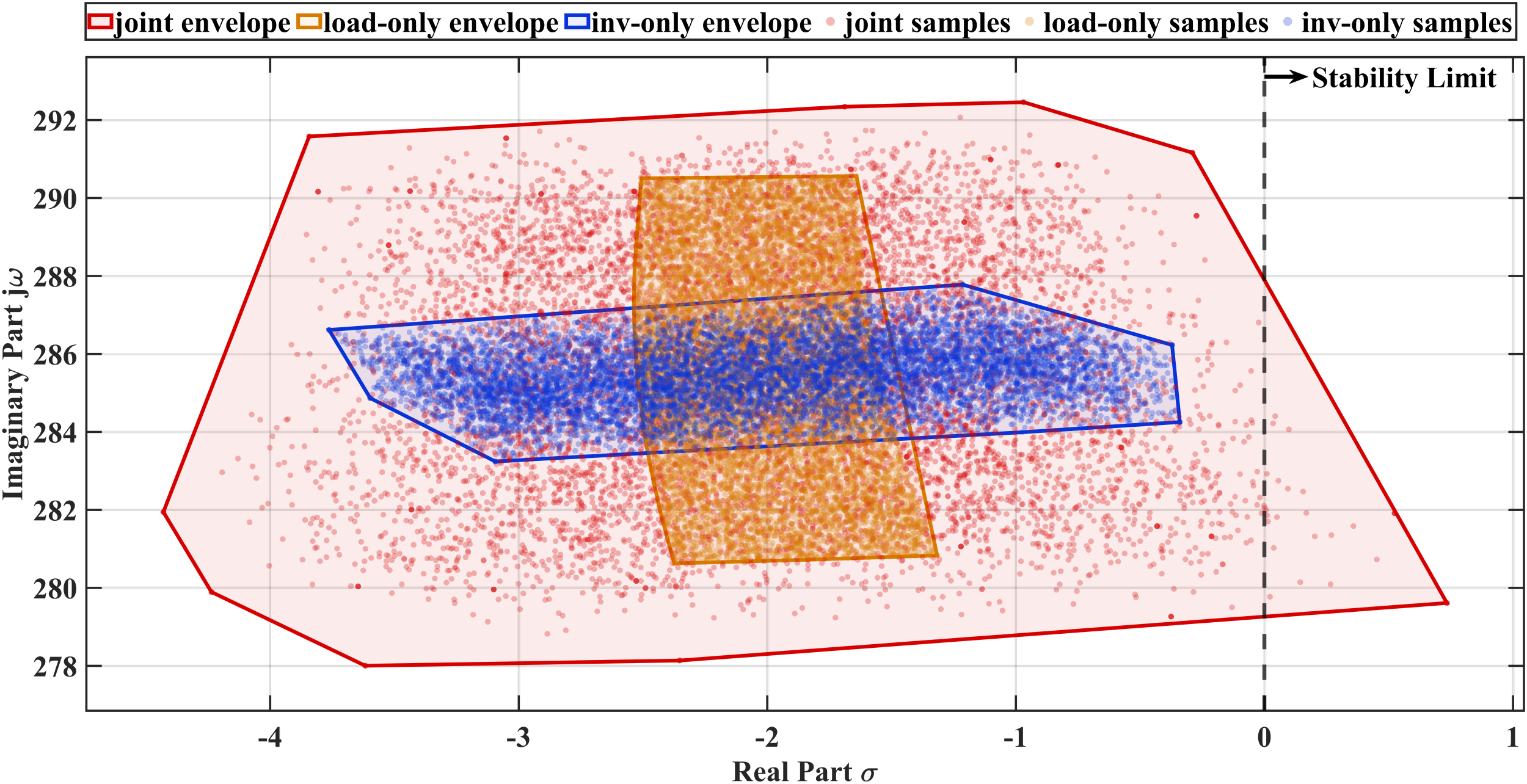}
    \caption{Attack reachable domains of the oscillatory modes under three attack scenarios.}
    \label{fig:ard_three_cases}
\end{figure}

\begin{figure}[t]
    \centering
    \includegraphics[width=0.45\textwidth]{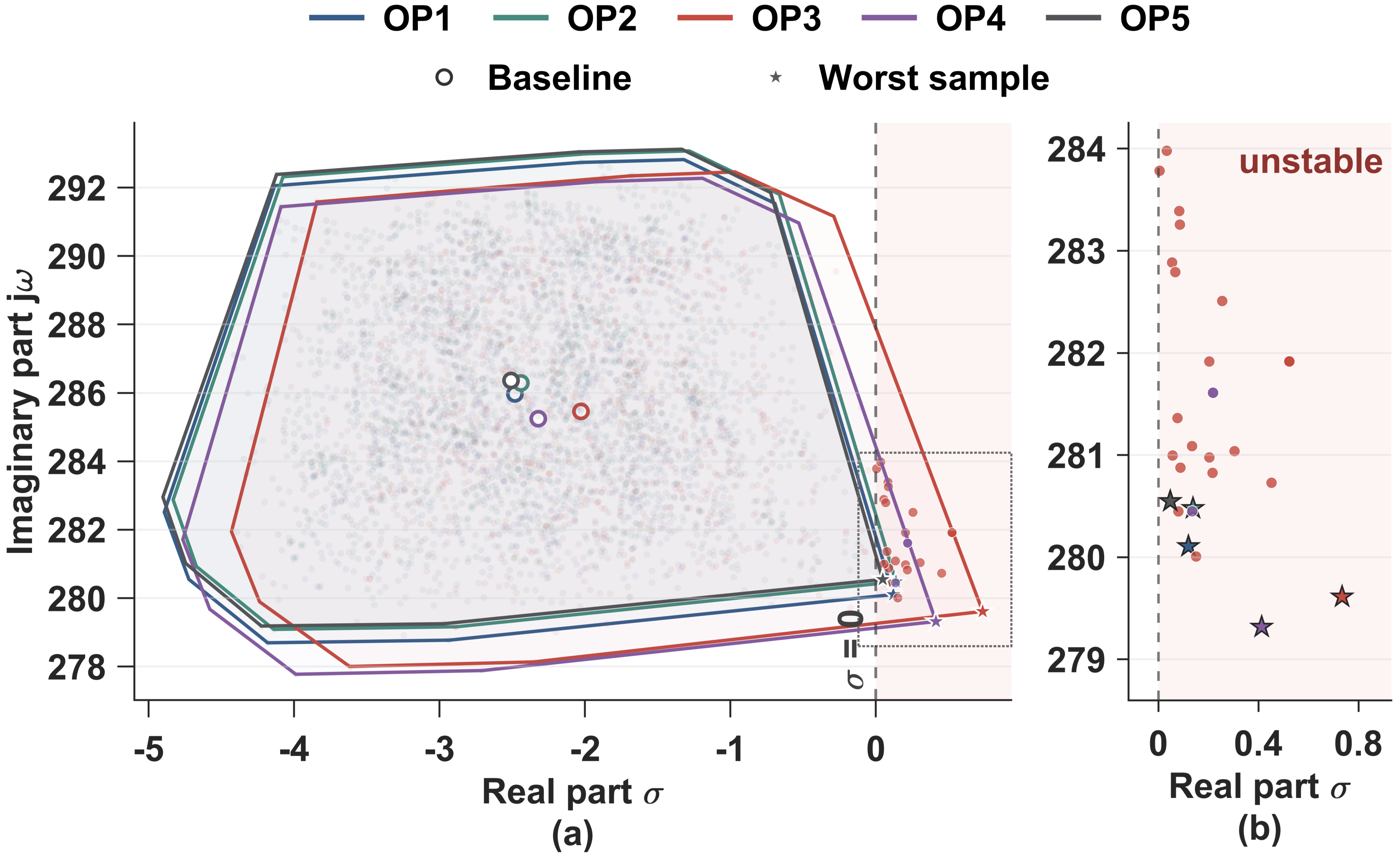}
    \caption{Attack reachable domains under coordinated attacks for five joint uncertainty realizations at the same time: (a) complete sample distribution, (b) enlarged view near the stability boundary.}
    \label{fig:ard_multi_ops}
\end{figure}

\vspace{-10pt}
\subsection{Validation of the Proposed Framework}
This subsection evaluates the effects of coordinated attacks relative to single-sided attacks and identifies uncertainty-aware attack time windows under joint PV and AIDC load uncertainty.
\subsubsection{Effects of Attack Coordination}
To evaluate the effect of attack coordination, the ARDs are compared under three scenarios: 1) an inverter-side-only scenario comprising inverter parameter tampering, 2) a load-side-only scenario comprising AI-induced AIDC demand variations, and 3) a coordinated scenario comprising both attack components.

In the inverter-side-only scenario, the five inverter control parameters listed in Table~\ref{tab:setup} were modified within relative bounds of 5\%, 5\%, 5\%, 10\%, and 10\%, respectively, while the load condition was held fixed. Table~\ref{tab:aidc_power_impact} presents illustrative cases with $A_w=13.12$, the maximum token-consumption amplification reported for a RAG inference-cost attack \cite{liu2026inference}. In the load-side-only scenario, we conservatively assume AIDC loads fluctuate within $15\%$ of their baseline values.

Fig.~\ref{fig:ard_three_cases} shows the ARDs at the evaluated representative time instant. Under the inverter-only
and load-only attacks, the rightmost eigenvalue real parts are $-0.341$ and $-1.317$, respectively. The ARDs remain in the stable region. Thus, at this operating point and within the imposed attack bounds, neither inverter parameter tampering nor AI-induced demand variation alone reaches an unstable eigenvalue location. By contrast, the coordinated ARD crosses the stability boundary, and the rightmost eigenvalue real part is $0.734$, which shows that the two attacks have a cooperative effect.

To quantify the variability of coordinated attack vulnerability under joint uncertainty, Fig.~\ref{fig:ard_multi_ops} extends the analysis in Fig.~\ref{fig:ard_three_cases} by comparing the coordinated attack ARDs across five representative joint uncertainty realizations at the same time instant. As shown in Fig.~\ref{fig:ard_multi_ops}(a), the sampled eigenvalue distributions and ARD envelopes vary substantially across the realizations. The open circles denote the baseline eigenvalues for the respective realizations, whereas the stars identify the samples with the largest eigenvalue real parts within each ARD. The enlarged view in Fig.~\ref{fig:ard_multi_ops}(b) further reveals substantial variation in the extent to which the ARDs enter the right half plane. These results indicate that a single deterministic realization cannot adequately characterize the coordinated attack vulnerability at a given time instant.

\subsubsection{Uncertainty-Aware Vulnerability Assessment}
\begin{figure*}[t]
    \centering
    \includegraphics[width=0.85\textwidth]{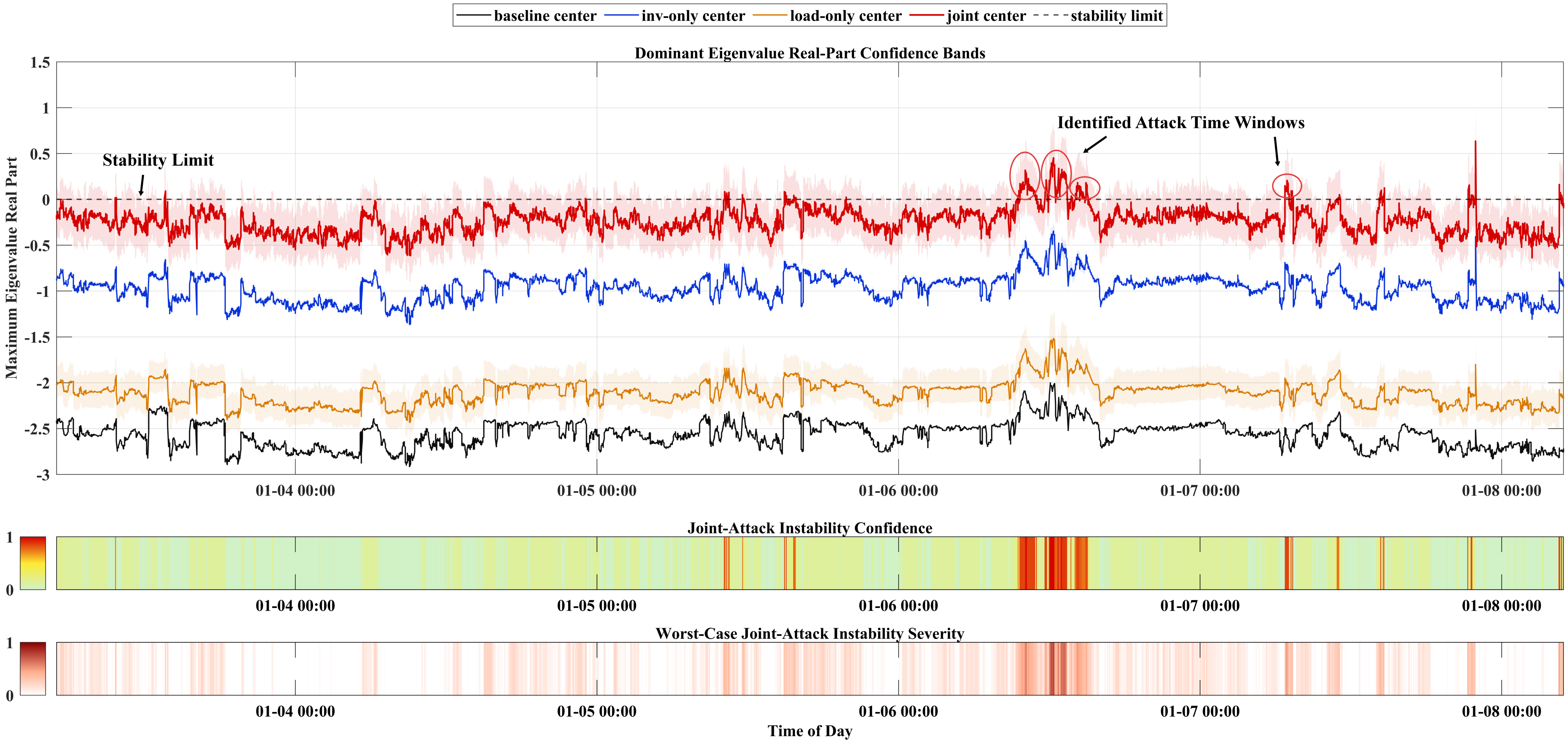}
    \caption{Five-day uncertainty-aware vulnerability assessment under coordinated attacks.}
    \label{fig:window_band}
\end{figure*}

The time-varying vulnerability of the AIDC microgrid was evaluated using the maximum real part of the closed-loop eigenvalues under the baseline, inverter-side-only, load-side-only, and coordinated attack scenarios. The coordinated attack scenario included both PV operating-point uncertainty and AI-induced demand response uncertainty.

Fig.~\ref{fig:window_band} reports the vulnerability assessment over 7,200 one-minute time instants. The maximum eigenvalue real parts of the baseline, inverter-only, and load-only cases remained below zero. By contrast, the coordinated attack trajectory approaches the stability boundary during several localized intervals. Although at least one joint uncertainty realization was unstable at 4,726 time instants ($65.64\%$), only 41 instants ($0.569\%$) satisfied $A_{t+\tau}\ge0.9$. These instants formed eight continuous attack windows. The confidence-weighted vulnerability was therefore concentrated within a small number of short intervals rather than distributed uniformly over the evaluated horizon.

The shaded band around the coordinated attack trajectory quantifies the variation induced by the joint uncertainty. At a given future time instant, different realizations of the PV operating point and AI-induced demand response can yield different stability classifications. Consequently, a single deterministic trajectory does not represent the range of evaluated outcomes. Therefore, uncertainty-aware assessment is necessary to avoid both optimistic and overly conservative conclusions.

The middle heatmap in Fig.~\ref{fig:window_band} reports the confidence-weighted attackability score. Elevated scores occur within several short intervals, indicating that feasible coordinated attacks reach unstable eigenvalue locations across a larger confidence weight of uncertainty realizations during these periods. The lower heatmap reports the joint-attack instability severity index. Regions with larger index values are sparse and largely coincide with intervals of elevated attackability. This concurrence identifies intervals associated with both prevalent attackability across the evaluated realizations and greater instability severity.

Overall, the results indicate that the proposed method extracts uncertainty-aware attack time windows from the five-day operating trajectory and distinguishes these intervals from the remaining evaluated horizon. Within the identified intervals, coordinated attacks exhibit both elevated confidence-weighted attackability and greater instability severity index values.

\subsubsection{Uncertainty-Aware Attack Vector Identification}
\label{subsubsec:ua_attack_vector_case}
After identifying the uncertainty-aware attack time windows, coordinated attack vectors that satisfy the prescribed effectiveness criterion within each selected window were sought. Whereas window identification determines when the system is vulnerable, vector identification determines which coordinated attacks retain destabilizing capability during each vulnerable interval.

For each identified window, candidate coordinated attack vectors are sampled from the feasible attack set, with each vector comprising inverter-side tampering and AIDC demand response components. A candidate is retained when its minimum confidence-weighted effectiveness over the window satisfied the threshold $\beta$ in \eqref{eq:ua_attack_vector_condition_conf}. Thus, the selected vector satisfies a window-wide uncertainty-aware criterion rather than maximizing instability at a single time instant.

Table~\ref{tab:universal_attack_vectors} lists coordinated attack vectors for three selected uncertainty-aware attack time windows. Their minimum confidence-weighted effectiveness scores are $0.98$, $0.98$, and $1.00$, all exceeding $\beta=0.9$. The corresponding maximum eigenvalue real parts are $0.839$, $0.776$, and $1.029$. It is observed that most of the identified vulnerable windows occur during periods with stronger source-load operating point variations. For example, the top two windows appear around noon on 2020-01-06, when the PV operating point is more sensitive to irradiance changes and the AIDC demand response can further reshape the load-side impedance. Their combined effect creates operating periods in which a fixed coordinated attack vector can remain effective under joint uncertainty.


\begin{table*}[!t]
\centering
\footnotesize
\caption{Identified window-specific coordinated attack vectors.}
\label{tab:universal_attack_vectors}
\scalebox{0.9}{
\begin{tabular}{c| c| c| c| c| c| c| c}
\hline
\textbf{Rank} 
& \textbf{Time} 
& \(\boldsymbol{k_{pv}}\) 
& \(\boldsymbol{k_{iv}}\) 
& \(\boldsymbol{k_{pi}}\) 
& \(\boldsymbol{J}\) 
& \(\boldsymbol{D_p}\) 
& \textbf{Load Modification} 
\\
\hline
1 
& 2020-01-06 12:15--12:25
& 0.411 (-5\%)
& 214.18 (+5\%)
& 0.499 (-5\%)
& 0.128 (+10\%)
& 32573 (-10\%)
& 
\begin{tabular}{c}
\(P_{\mathrm{dc}}\in[20.644,23.224]~\mathrm{kW}\) \\
\(Q_{\mathrm{dc}}\in[26.924,29.372]~\mathrm{kvar}\)
\end{tabular}
\\

2 
& 2020-01-06 12:05--12:10
& 0.411 (-5\%)
& 214.18 (+5\%)
& 0.499 (-5\%)
& 0.128 (+10\%)
& 32573 (-10\%)
& 
\begin{tabular}{c}
\(P_{\mathrm{dc}}\in[20.277,22.811]~\mathrm{kW}\) \\
\(Q_{\mathrm{dc}}\in[25.375,27.682]~\mathrm{kvar}\)
\end{tabular}
\\

3 
& 2020-01-07 21:55--22:00
& 0.411 (-5\%)
& 214.18 (+5\%)
& 0.499 (-5\%)
& 0.104 (-10\%)
& 32573 (-10\%)
& 
\begin{tabular}{c}
\(P_{\mathrm{dc}}\in[21.888,24.624]~\mathrm{kW}\) \\
\(Q_{\mathrm{dc}}\in[34.621,37.769]~\mathrm{kvar}\)
\end{tabular}
\\
\hline
\end{tabular}
}
\end{table*}

\begin{figure}[t]
    \centering
    \includegraphics[width=0.4\textwidth]{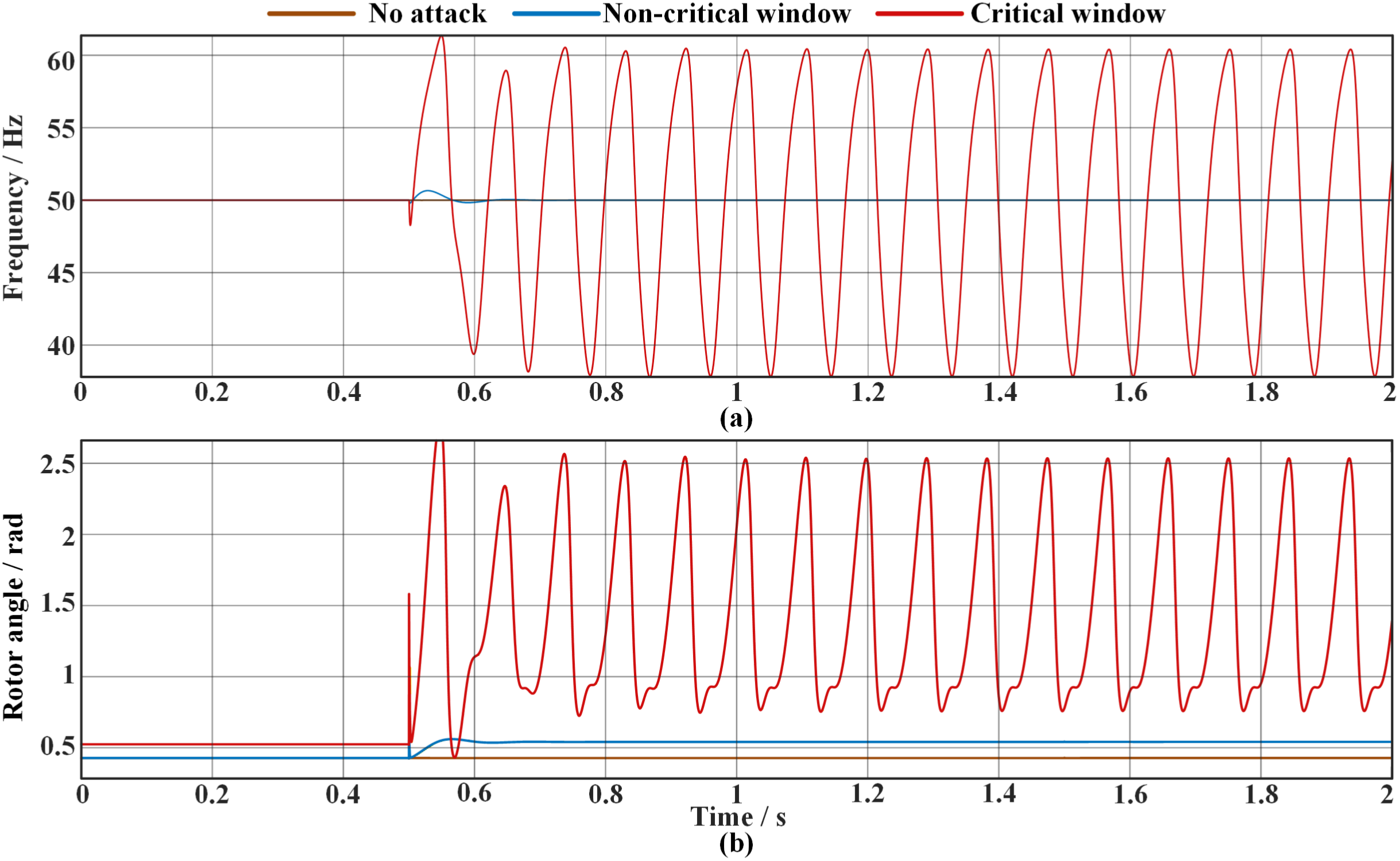}
    \caption{Time-domain responses to coordinated attacks launched in different time windows: (a) PV-inverter frequency, (b) VSG rotor angle.}
    \label{fig:impact_analysis}
\end{figure}

\subsubsection{Time Domain Validation of Critical Attack Time Windows}
To verify the physical impact of the coordinated attack under different time windows, we compare the time-domain responses under three scenarios: no attack, a coordinated attack launched outside the identified critical time window, and a coordinated attack launched within the critical time window. The inverter frequency and virtual synchronous generator (VSG) rotor angle responses are shown in Fig.~\ref{fig:impact_analysis}.

Fig.~\ref{fig:impact_analysis}(a) and (b) show the PV inverter frequency and VSG rotor angle responses, respectively. Without attacks, both responses remain bounded. In the evaluated noncritical-window case, the inverter frequency remained between 49.79 and 50.65~Hz. In the critical-window case, the frequency ranged from 37.80 to 61.35~Hz, the PV inverter frequency exhibits sustained oscillations with peak absolute deviations exceeding 20\% of the nominal frequency, accompanied by sustained oscillations in the VSG rotor angle. These results shows that the destabilizing effect of coordinated attacks is time-dependent and that the identified critical time windows correspond to actual dynamic instability risks.


\vspace{-12pt}
\section{Conclusion}

In this paper, we propose an uncertainty-aware framework for assessing AIDC microgrid vulnerability under computing-power coordinated attacks. First, the framework links adversarial AI requests to changes in AIDC power demand and impedance, then combines these responses with confidence-weighted PV uncertainty to construct ARDs of critical closed-loop eigenvalues. Afterward, for each joint uncertainty realization, the framework determines whether any feasible coordinated attack can move the critical closed-loop eigenvalue into the right-half plane and aggregates these outcomes using confidence weights. Finally, the resulting attackability score quantifies the prevalence of attackable conditions across plausible future realizations, enabling the identification of attack time windows and effective attack vectors. Case studies show that AIDC impedance varies across load conditions and that coordinated attacks can produce instability unattainable by either attack component alone. Over a five-day trajectory, high-confidence vulnerability is concentrated in sparse intervals, with fixed coordinated attack vectors remaining effective over selected windows. The results of the electromagnetic transient simulation further show bounded responses outside these windows but sustained frequency oscillations and rotor angle deviations within them. These findings show that AIDC microgrid vulnerability is time-dependent and shaped by cross-domain coupling, providing a basis for targeted detection and protection during credible high-risk periods.

\vspace{-10pt}
\bibliographystyle{IEEEtran}

\bibliography{mybibfile}

\end{spacing}

\end{document}